\documentclass[aps,amsmath,amssymb,twocolumn,nofootinbib,pra,superscriptaddress,longbibliography]{revtex4-2} 

\usepackage[colorlinks,citecolor=blue,urlcolor=blue,bookmarks=false,hypertexnames=true]{hyperref} 

\usepackage{graphicx}
\usepackage{subfigure}

\usepackage{amsmath,amsfonts,bm,thmtools}
\usepackage{amsmath}            
\usepackage{amssymb}

\newcommand{ \tr }{ \textrm{Tr} }

\newcommand{\Symbol}[2]{{#1}_{\textrm{#2}}}

\begin{document}

\title{Non-Markovian skin effect}

\author{Po-Chen Kuo}
\email{These authors contributed equally to this work.}
\affiliation{Department of Physics, National Cheng Kung University, 701 Tainan, Taiwan}
\affiliation{Center for Quantum Frontiers of Research and Technology, NCKU, 701 Tainan, Taiwan}
\affiliation{Theoretical Quantum Physics Laboratory, Cluster for Pioneering Research, RIKEN, Wakoshi, Saitama 351-0198, Japan}
\author{Shen-Liang Yang}
\email{These authors contributed equally to this work.}
\affiliation{Department of Physics, National Cheng Kung University, 701 Tainan, Taiwan}
\affiliation{Center for Quantum Frontiers of Research and Technology, NCKU, 701 Tainan, Taiwan}
\author{Neill Lambert}
\email{nwlambert@gmail.com}
\affiliation{Theoretical Quantum Physics Laboratory, Cluster for Pioneering Research, RIKEN, Wakoshi, Saitama 351-0198, Japan}
\affiliation{Center for Quantum Computing, RIKEN, Wakoshi, Saitama 351-0198, Japan}
\author{Jhen-Dong Lin}
\affiliation{Department of Physics, National Cheng Kung University, 701 Tainan, Taiwan}
\affiliation{Center for Quantum Frontiers of Research and Technology, NCKU, 701 Tainan, Taiwan}
\author{Yi-Te Huang}
\affiliation{Department of Physics, National Cheng Kung University, 701 Tainan, Taiwan}
\affiliation{Center for Quantum Frontiers of Research and Technology, NCKU, 701 Tainan, Taiwan}
\author{Franco Nori}
\affiliation{Theoretical Quantum Physics Laboratory, Cluster for Pioneering Research, RIKEN, Wakoshi, Saitama 351-0198, Japan}
\affiliation{Center for Quantum Computing, RIKEN, Wakoshi, Saitama 351-0198, Japan}
\affiliation{Physics Department, The University of Michigan, Ann Arbor, Michigan 48109-1040, USA.}
\author{Yueh-Nan Chen}
\email{yuehnan@mail.ncku.edu.tw}
\affiliation{Department of Physics, National Cheng Kung University, 701 Tainan, Taiwan}
\affiliation{Center for Quantum Frontiers of Research and Technology, NCKU, 701 Tainan, Taiwan}
\affiliation{Physics Division, National Center for Theoretical Sciences, Taipei 10617, Taiwan}


\begin{abstract}

The Liouvillian skin effect and the non-Hermitian skin effect have both been used to explain the localization of eigenmodes near system boundaries, though the former is arguably more accurate in some regimes due to its incorporation of quantum jumps.
However, these frameworks predominantly focus on weak Markovian interactions, neglecting the potentially crucial role of memory effects. To address this, we investigate, utilizing the powerful hierarchical equations of motion method,  how a non-Markovian environment can modify the Liouvillian skin effect. We demonstrate that a non-Markovian environment can induce a ``thick skin effect", where the skin mode broadens and shifts into the bulk. {We further identify that the skin-mode quantum coherence can only be generated when the coupling contains counter-rotating terms}, leading to the coherence-delocalization and oscillatory relaxation with a characteristic linear scaling with system size. Remarkably, both the skin-mode and steady-state coherence exhibit resistance to decoherence from additional environmental noise. These findings highlight the profound impact of system-bath correlations on relaxation and localization, revealing unique phenomena beyond conventional Markovian approximations.
\end{abstract}
\maketitle

\emph{Introduction}---The skin effect, where wavefunctions localize at system boundaries under specific conditions, is crucial in condensed matter physics~\cite{Lee2019,Masatoshi2021,Lin2023} and quantum optics~\cite{Linhu2020}. Extensive research has explored the non-Hermitian skin effect~\cite{Zhang2021,Taylor2021,2Zhang2021}, where non-trivial skin-topological modes arise from gain and loss interplay~\cite{YongChun2022}, impacting topological classification~\cite{LiuTao2020,Hofmann2020,Palacios2021,nakai2024topological,Leefmans2024}, dynamical phase transitions~\cite{Moon2021,Shinsei2023}, and quantum state engineering~\cite{Zhong2019}.

Recent theoretical advances have extended the skin effect concept to Lindblad master equations, termed the Liouvillian skin effect (LSE)~\cite{Haga2021,Fan2022,Hamanaka2023,Clerk2023,Wang2023}. This generalization emerges from the Liouville-von Neumann equation~\cite{HeinzPeter2009}, governing the temporal evolution of the system density matrix $\rho_{\text{s}}(t)$. Current LSE investigations predominantly rely on the Born-Markov approximation~\cite{supplemental}. The LSE, incorporating quantum jumps from system-environment interactions, offers a more accurate description of open quantum systems~\cite{Fabrizio2018,Fabrizio2019,pochen2020,Fabrizio2020}. This phenomenon manifests as anomalous localization of eigenmodes at system boundaries, driven by nonreciprocal dissipation, potentially extending relaxation time $\tau$ without closing the Liouvillian gap $\Delta$~\cite{Haga2021,Wang2023}.

\begin{figure*}[]
	\centering
    \includegraphics[width = 2\columnwidth]{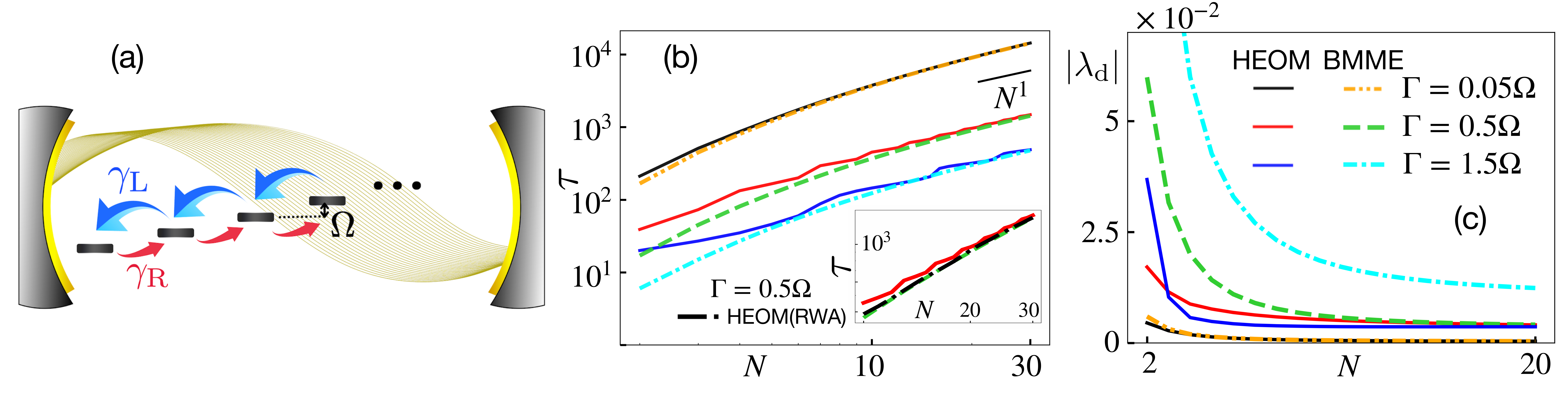}
	\caption{(a) Schematic of an N-site 1D chain with detuning $\Omega$ and non-reciprocal hopping ($\gamma_{\text{L}}\neq\gamma_{\text{R}}$) coupled to a bosonic environment. (b) Relaxation time $\tau$ vs. $N$ in Markovian (dashed) and non-Markovian (solid) regimes for $\Gamma=0.05, 0.5, 1.5\Omega$. Black line shows $N^1$ scaling. Inset: $\tau$ vs. $N$ for BMME, HEOM, and RWA-HEOM at $\Gamma=0.5\Omega$. (c) Dominant eigenvalue magnitude $|\lambda_{\text{d}}|$ vs. $N$ for different $\Gamma$. Parameters: $T=1.44\Omega$, $W=\Omega$, $l_{\text{max}}=20$, $m_{\text{max}}=4$ (8 for $\Gamma=1.5\Omega$).}
	\label{fig1}
\end{figure*}

While Markovian approaches have provided valuable insights, many realistic scenarios involve strong system-environment interactions or long-lasting correlations, leading to non-Markovian effects with profound implications~\cite{HeinzPeter2009,HeinzPeter2011,Gribben2022}. These effects enhance quantum coherence~\cite{Mang2010}, protect quantum information~\cite{HeinzPeter2016}, enable quantum metrology~\cite{Plenio2012,Diego2020}, and modify quantum transport~\cite{Flindt2008,ZhengXiao2009,YuehNan2009,Moreira2020,Huang2023}. They can also improve quantum device performance~\cite{Hill2022} and influence chemical reaction kinetics~\cite{Gurin2012}.

An intriguing aspect to consider is how non-Markovian dynamics impacts the LSE, as current studies primarily rely on the Born-Markov approximation. This assumption breaks down for ultrastrong system-bath coupling~\cite{Geiser2012,Stassi2013,GuanTing2018,Anton2019,Lambert2019} and in many realistic settings involving complex environments with memory effects~\cite{Vega2017,Andersson2019,Kuo2023}. Therefore, a comprehensive understanding of the skin effect in the non-Markovian regime is needed. A central question is whether the established relationship between relaxation time $\tau$ and system size $N$, expressed in terms of the localization length $\xi$, remains valid~\cite{Haga2021} under non-Markovian conditions:
\begin{equation}
\tau\overset{?}{\sim}\frac{1}{\Delta} + \frac{N}{\Delta \xi}.
\end{equation}
In this work, we employ the Hierarchical equations of motion (HEOM) approach, offering non-perturbative~\cite{Yan2012,lambert2020bofinheom,Huang2023} and non-Markovian characterization~\cite{Tanimura06,Tanimura02} of environmental effects, to investigate the non-Markovian skin effect. Our abstract model reveals novel features including: (i) a ``thick skin" effect with delocalized steady-state population at the boundary, {(ii) emergence of persistent cross-site quantum coherence in skin-localized modes under decoherence noise, and (iii) secondary skin-localized modes modulating oscillatory relaxation dynamics.}

We here consider a single particle in an $N$-site 1D lattice coupled to a bosonic bath as shown in Fig.~\ref{fig1}. The system Hamiltonian is given by (with $\hbar=1$ throughout) $H_{\text{s}}=\sum_{n=1}^{N}\omega_n d^\dagger_n d_n $, where $d_n$ ($d_n^{\dagger}$) corresponds to the creation (annihilation) operator of the particle within the $n$th site. We assume identical energy detuning $\Omega=\omega_n-\omega_{n-1}$ between adjacent sites, with the particle initially in the highest energy state.. To induce the skin effect, we introduce thermally-induced nonreciprocal hopping via a bosonic bath with correlation function
\begin{equation}
\begin{aligned}
C(\tau)
=\frac{1}{2\pi}\int_{0}^{\infty} d\omega 
J(\omega)\Big\{n^{\textrm{eq}}(\omega)e^{i\omega \tau}
+[n^{\textrm{eq}}(\omega)+1]e^{-i\omega \tau}
\Big\},
\label{eq:C_b}
\end{aligned}
\end{equation}
where $n^{\textrm{eq}}(\omega)=\{\exp[\omega/k_{\textrm{B}}T]-1\}^{-1}$, representing the Bose–Einstein distribution with $k_{\textrm{B}}$ as the Boltzmann constant and $T$ as the temperature. We assume a Drude-Lorentz spectral density $J(\omega)=4\Gamma W\omega/(\omega^2+W^2)$, where $\Gamma$ represents the coupling strength between the system and the bosonic reservoir with bandwidth $W$. {We also explore the non-Markovian skin effect under other choices of bath spectral densities to understand the role of the environment's structure \cite{supplemental}.} 
Assuming closely spaced sites enabling collective coupling to a common bath~\cite{Tanimura04,Tanimura03}, the system-bath interaction is given by $H_{\text{I}} = V_{\Symbol{\sigma}{b}}\sum_{k}g_k(a_k + a_k^{\dagger})$, where system interaction operator, $V_{\Symbol{\sigma}{b}} = \sum_{n}^{N} (d_{n}^{\dagger}d_{n+1} + d_{n+1}^{\dagger}d_{n})$, mediates nearest-neighbor hopping. For distant sites coupled to independent baths, $H_{\text{I}}=\sum_{n}V_{\Symbol{\sigma}{b},n}\sum_{k}g_{n,k}(a_{n,k} + a_{n,k}^{\dagger})$ with $V_{\Symbol{\sigma}{b},n}=d_{n}^{\dagger}d_{n+1}+d_{n+1}^{\dagger}d_{n}$~\cite{Tanimura01}. While both cases can exhibit the LSE, we focus on the richer collective coupling scenario, with results for the non-collective case presented in the supplemental material~\cite{supplemental}.

For the Markovian case, nonreciprocal hopping can be determined by the asymmetric dissipation rate for left and right hopping, $\gamma_{\text{L}}=J(\Omega)[n^{\text{eq}}(\Omega)+1]$m and  $\gamma_{\text{R}}=J(\Omega)n^{\text{eq}}(\Omega)$, respectively,  encoded in the Lindblad operators $L_{\text{L}}=\sqrt{\gamma_{\text{L}}}\sum_{n=1}^{N-1}d_{n}^{\dagger}d_{n+1}$ and $L_{\text{R}}=\sqrt{\gamma_{\text{R}}}\sum_{n=1}^{N-1}d_{n+1}^{\dagger}d_{n}$, according to BMME in Eq.~(A23)~\cite{supplemental}. This thermally-induced asymmetry allows for flexible control of the $\gamma_{\text{L}}/\gamma_{\text{R}}$ ratio by adjusting the temperature. This asymmetric thermally-induced hopping is realizable in various systems, including photonic topological lattices~\cite{Leefmans2022,Parto2023}, ultracold atoms in an optical lattice driven by laser-assisted hopping with spontaneous emission~\cite{Hofstadter2013,Harper2013,Xiang2023}, trapped-ion systems where lasers couple internal states to external motional sideband states~\cite{Hamann1998,Kaufman2012,Zohar2022}, and 1D quantum dots where transport electrons couple to photon bath~\cite{Lu2007,Brandes2008,Mishchenko2015,Jiajun2020,KondoQED2023}. For simplicity, we set $\gamma_{\text{L}}/\gamma_{\text{R}}\approx2 $ throughout.

In the non-Markovian case, nonreciprocal hopping necessitates going beyond the Born-Markov and secular approximations, incorporating non-Markovian and non-perturbative effects~\cite{Tanimura08,Tanimura06}. This is achieved using the interaction operator $V_{\Symbol{\sigma}{b}}$ within the HEOM generator~\cite{Yan2012,Tanimura05,Huang2023} in Eq.~(\ref{eq:HEOML}), as detailed in the following section.

\emph{HEOM-derived skin effect}---To incorporate the non-Markovian and non-perturbative effects we employ the HEOM method, which relies on a Liouvillian superoperator operating on a system's (s) space expanded by \textit{auxiliary density operators} (ADOs). These ADOs encode environmental influence and memory effects arising from system-bath interactions~\cite{Tanimura07}. Each ADO relates to an exponential term in the decomposition of the correlation function $C(\tau)=\sum_{u=\mathbb{R},\mathbb{I}}(\delta_{u,\mathbb{R}}+i\delta_{u,\mathbb{I}})C^{u}(\tau)$, where $\mathbb{R}$ and $\mathbb{I}$ respectively denote the real and imaginary parts of $C^{u}(\tau)=\sum_{l}^{l_{\text{max}}}\xi_{l}^{u}\exp(-\chi_{l}^{u}\tau)$~\cite{supplemental}. The HEOM generator $\hat{\mathcal{L}}_{\text{H}}$, operating on the combined system-ADO space $\rho_{\text{s}+\text{ADO}}(t)$, characterizes the dynamics as follows~\cite{Mauro2022,Huang2023,debecker2024,JhenDong2024}
\begin{equation}
\begin{aligned}
{\frac{\partial}{\partial_{t}}\textbf{vec}[\rho_{\text{s}+\text{ADO}}(t)]=\hat{\mathcal{L}}_{\text{H}}\textbf{vec}[\rho_{\text{s}+\text{ADO}}(t)],}
\label{eq:HEOML}
\end{aligned}
\end{equation}%
{where $\textbf{vec}[\cdot]$ denotes vectorization for eigensolving.} This HEOM generator encodes system-bath interactions through $V_{\Symbol{\sigma}{b}}$. Subsequently, the right and left eigenmodes of the $\hat{\mathcal{L}}_{\text{H}}$ are defined by~\cite{JhenDong2024}
\begin{figure}[]
	\centering
    \includegraphics[width = 1.\columnwidth]{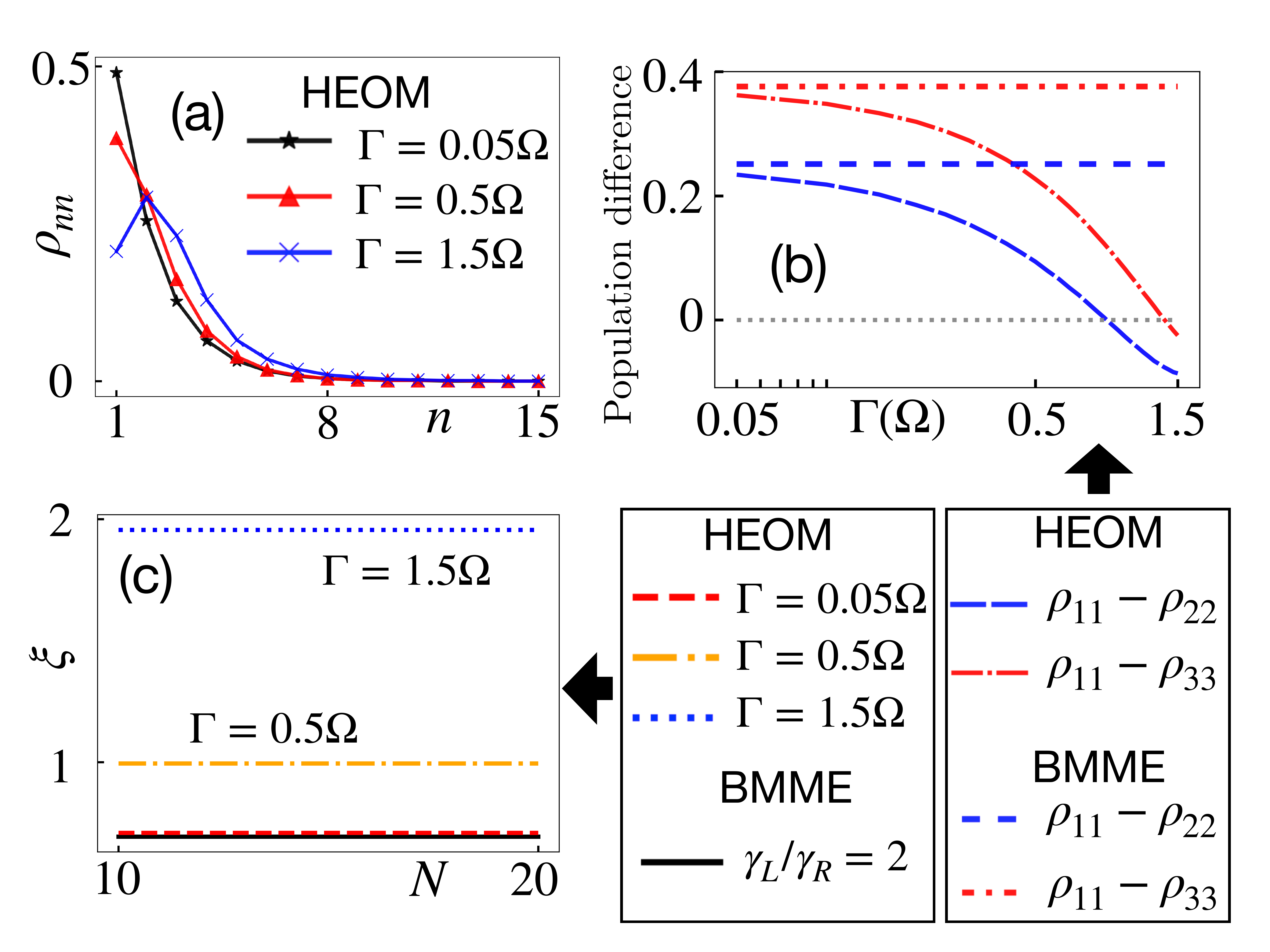}
\caption{(a) Site populations in $\pmb{\rho}^{\text{R}}_{0,\text{s}}$ for different $\Gamma$ in the non-Markovian regime (HEOM). (b) Population differences $\rho_{11}-\rho_{22}$ (dashed) and $\rho_{11}-\rho_{33}$ (dash-dotted) vs. $\Gamma$ for BMME and HEOM. (c) Localization length $\xi$ of $\hat{\rho}_{\text{d}}$ vs. $N$ for different $\Gamma$ in Markovian (BMME) and Non-Markovian (HEOM) regimes. Here, $\gamma_L/\gamma_R = 2$ for BMME.}\label{fig2}
\end{figure}
\begin{equation}\label{eq:eigenmode}
\hat{\mathcal{L}}_{\text{H}}\pmb{\rho}^{\text{R}}_{i}=\lambda_{i}\pmb{\rho}^{\text{R}}_{i}, \:~~~ \pmb{\rho}^{\text{L}\dagger}_{j}\hat{\mathcal{L}}_{\text{H}}=\lambda_{j}\pmb{\rho}^{\text{L}\dagger}_{j},
\end{equation}
 where $\lambda_{i(j)}$ is the $i$th ($j$th) eigenvalue with corresponding right $\pmb{\rho}^{\text{R}}_{i}$ (left $\pmb{\rho}^{\text{L}}_{j}$) eigenmode.
 The HEOM-space inner product of two ADO vectors is defined as $(\pmb{\rho}_{a}^{v}|\pmb{\rho}_{b}^{v'})=\sum_{k}\rho_{a}^{v\dagger}[k]\cdot\rho_{b}^{v'}[k]$, where $k$ indexes elements in each eigenmode. The trace norm of an ADO is then $||\pmb{\rho}_{i}^{v}|| = (\sum_{k}\pmb{\rho}_{i}^{v\dagger}[k]\pmb{\rho}_{i}^{v}[k])^{1/2}$. {The first right eigenmode $\pmb{\rho}^{\text{R}}_0$ with $\lambda_0=0$ corresponds to the stationary state $\hat{\pmb{\rho}}_{\text{ss}}=\pmb{\rho}^{\text{R}}_0/ \tr (\pmb{\rho}^{\text{R}}_{0,\text{s}})$, ensuring probability preservation. Here, the subscript s in $\pmb{\rho}^{\text{R}}_{0,\text{s}}$ denotes the system reduced density matrix within the ADOs.} The HEOM dynamics can be decomposed into the right eigenmodes~\cite{supplemental}:
\begin{equation}\label{eq:dynamics}
    {\textbf{vec}[\rho_{\text{s}+\text{ADO}}(t)]}=\hat{\pmb{\rho}}_{\text{ss}}+\sum_{i=1}^{N_{\text{tot}}}c_ie^{\lambda_i t}\hat{\pmb{\rho}}^{\text{R}}_{i},
\end{equation}
{where $N_{\text{tot}} = N_{\text{s}}^2 \times N_{\text{ADO}}$, with $N_{\text{s}}$ and $N_{\text{ADO}}$ denoting the dimension of $H_{\text{s}}$ and the number of ADO, respectively.} Here, $c_{i}=||\pmb{\rho}^{\text{L}}_i||||\pmb{\rho}^{\text{R}}_i|| (\hat{\pmb{\rho}}^{\text{L}}_i|\textbf{vec}[\hat{\pmb{\rho}}_{\text{ini}}])$, where $\hat{\pmb{\rho}}_{\text{ini}}$ represents an arbitrary initial state. We assume that $\hat{\pmb{\rho}}_{\text{ini},\text{s}}=|N\rangle\langle N|$ with no initial system-bath correlation. Following Ref.~\cite{Haga2021}, the long-time system dynamics are governed by the {steady-state eigenmode} $\hat{\pmb{\rho}}_{\text{ss}}$ and the dominant eigenmode $\hat{\pmb{\rho}}_{\text{d}}$ with eigenvalue $\lambda_{\text{d}}$. The relaxation time $\tau$, defined as the time at which the contribution of $\hat{\pmb{\rho}}_{\text{d}}$ surpasses all other eigenmodes, is when $|c_{\text{d}}(\tau)|=e^{-1}$, where $c_{\text{d}}(\tau)=c_{\text{d}}\exp{(-\tau\lambda_{\text{d}})}$. {While $\hat{\pmb{\rho}}_{\text{d}}$ is the dominant mode, virtual processes arising from counter-rotating terms can introduce secondary skin modes that also influence relaxation, unlike in the Markovian regime.} Furthermore, the HEOM-derived coefficient $c_{\text{d}}$ exhibits an $N$ dependence, following the relation $||\pmb{\rho}^{\text{L}}_i||||\pmb{\rho}^{\text{R}}_i||\sim e^{O(N/\xi_{\text{H}})}$, where $\xi_{\text{H}}$ is the HEOM-derived localization length. The HEOM-derived relaxation time is thus:
\begin{equation}\label{LSE}
\begin{aligned}
\tau\sim\frac{1}{\lambda_{\text{d}}} + \frac{N}{\lambda_{\text{d}} \xi_{\text{H}}}.
\end{aligned}
\end{equation}
We aim to demonstrate the hallmark of non-Markovian LSE by examining the relaxation time $\tau$ and the dominant eigenvalue$\lambda_{\text{d}}$, which corresponds to the Liouvillian gap in the Markovian limit.

{\emph{Relaxation time}---Both the Markovian (BMME) and non-Markovian (HEOM) approaches are employed here. We observe two key trends as shown in Fig.~\ref{fig1}. In the strong-coupling regime ($\Gamma\geq 0.5\Omega$), there is a notable difference between the results of BMME and HEOM. The HEOM yields a slower relaxation (larger $\tau$) corresponding to a larger dominant eigenvalue $\lambda_{\text{d}}$ [Fig. \ref{fig1}(b)] compared that of the BMME.}

{The impact of the non-Markovian effect is revealed by comparing HEOM and BMME results, particularly in the relaxation-determining parameter $\lambda_{\text{d}}$. For instance, when $\Gamma=0.05\Omega$, the values of $\lambda_{\text{d}}$ are consistent between BMME and HEOM under the Markovian condition. In contrast, when $\Gamma>0.5\Omega$, the deviation in $\lambda_{\text{d}}$ is obvious, indicating a transition to the non-Markovian regime.}

{Moreover, as $N$ increases, both approaches exhibit a direct linear relationship in $\tau$, while $\lambda_{\text{d}}$ [Fig. \ref{fig1}(c)] and the localization length $\xi_{\text{H}}$ [Fig. \ref{fig2}(c)] approach constant values. This signifies that even under the non-Markovian dynamics, the LSE persists, as inferred from Eq. (\ref{LSE}), and governs the long-time relaxation behavior at larger system sizes. Furthermore, in the strong-coupling regime, the relaxation time displays additional small oscillations. Notably, when we implement the rotating-wave approximation in HEOM (RWA-HEOM) to remove the counter-rotating terms ($d_{n+1}^{\dagger}d_n a_k^{\dagger}$ and $a_k d_n^{\dagger}d_{n+1}$ in $H_{\text{I}}$), these oscillations disappear. This indicates that not every non-Markovian effect can induce the oscillations. \textit{Only the virtual processes generated by the counter-rotating terms can induce such oscillations.} The connection between the oscillations and virtual processes will be further investigated later.}

{\emph{Thick skin effect}---A key signature of the Liouvillian skin effect is eigenmode localization. We focus on the steady-state eigenmode $\pmb{\rho}^{\text{R}}_{0,\text{s}}$ to examine the localization of the site population. In weak-coupling scenarios, this population is predominantly localized at the system's boundary. However, in the strong-coupling regime, non-Markovian dynamics can induce gradual delocalization of $\pmb{\rho}^{\text{R}}_{0,\text{s}}$, redistributing population across energy levels [Fig. \ref{fig2}(a)]. This redistribution stems from the enhanced system-bath interactions, facilitating energy and information exchange between the system and bath. Consequently, higher energy levels can maintain non-zero populations in the steady-state eigenmode, effectively ``repopulating" levels that typically deplete in a Markovian system. When $\Gamma\approx \Omega$, the population extends to the second or third site, potentially exceeding that of the first site. While this deeper and broadening localization of the population extends further into the chain, it still adheres to LSE principles. We term this the ``Thick skin effect", which cannot be captured by BMME, highlighting the inadequacy of Markovian approaches in this domain [Fig. \ref{fig2}(b)].} The ``Thick Skin Effect" manifests as an expansion in the localization length $\xi_{\text{H}}$ for $\hat{\pmb{\rho}}_{\text{d}}$ [Fig. \ref{fig2}(c)]. In the non-Markovian regime, this expansion adheres to LSE behavior and remains independent of system size $N$, consistent with LSE phenomena.

\begin{figure}[]
	\centering
    \includegraphics[width = 1.\columnwidth]{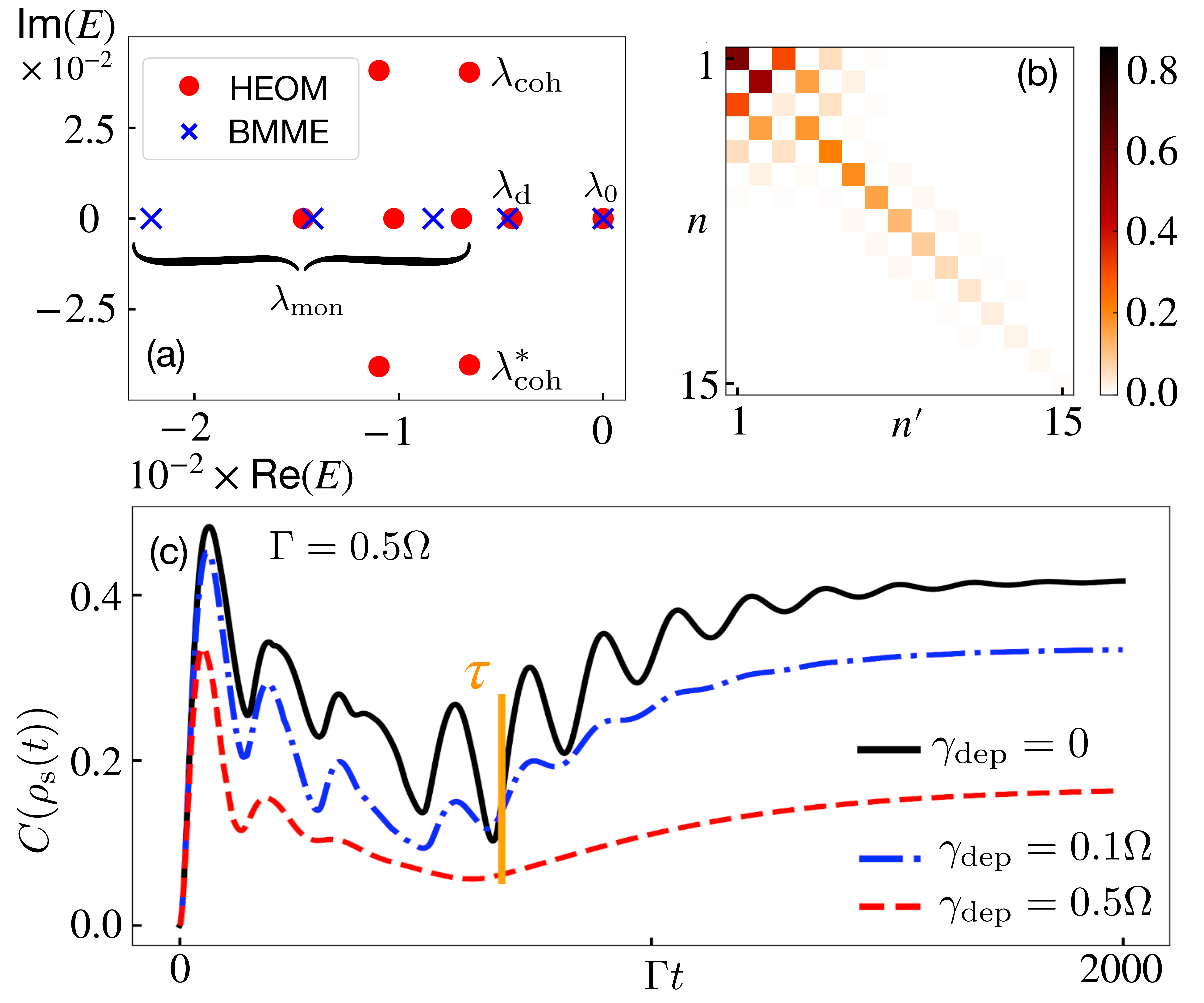}
	\caption{(a) Eigenspectrum obtained from BMME (blue) and HEOM (red) Liouvillian superoperators. Eigenvalues $\lambda_0$, $\lambda_{\text{d}}$, $\lambda_{\text{coh}}^{(\dagger)}$, and $\lambda_{\text{mon}}$ correspond to $\hat{\pmb{\rho}}_{\text{ss}}$, $\hat{\pmb{\rho}}_{\text{d}}$, $\hat{\pmb{\rho}}_{\text{coh}}^{(\dagger)}$, and $\hat{\pmb{\rho}}_{\text{mon}}$, respectively. (b) Color plots of the reduced density matrix extracted from $\hat{\pmb{\rho}}_{\text{d}}$ for $N=15$ obtained using HEOM. (c) Dynamics of coherence $C(\rho_{\text{s}}(t))$ at $\Gamma = 0.5\Omega$ under the dephasing applied to all sites with the rate $\gamma_{\text{dep}}$.} 
	\label{fig:4}
\end{figure}

{\emph{Emergence and resilience of quantum coherence induced by the virtual process}---The eigenspectrum of the HEOM generator reveals a rich structure that significantly deviates from the BMME Liouvillian, as illustrated in Fig. 3(a). The steady-state eigenmode $\hat{\pmb{\rho}}_{\text{ss}}$ corresponds to $\lambda_0$, while the primary skin mode $\hat{\pmb{\rho}}_{\text{d}}$ is associated with $\lambda_{\text{d}}$. Notably, the HEOM approach unveils secondary skin modes $\hat{\pmb{\rho}}_{\text{coh}}^{(\dagger)}$ with complex eigenvalues, $\lambda_{\text{coh}}^{(\dagger)}$, showing oscillatory behavior. The monotonically decaying modes $\hat{\pmb{\rho}}_{\text{mon}}$ are represented by $\lambda_{\text{mon}}$. Crucially, both $\hat{\pmb{\rho}}_{\text{d}}$ and $\hat{\pmb{\rho}}_{\text{mon}}$ in HEOM exhibit eigenvalues closer to $\lambda_0$ compared to that of BMME, accounting for the extended relaxation time.}

{A striking feature of the HEOM-derived eigenmodes, including the skin modes, is the presence of cross-site coherence [Fig. 3(b)], while the quantum coherence between any two adjacent sites is zero. Such a coherence is absent in both BMME and RWA-HEOM approaches. This coherence, quantified by the $l_1$ norm $C(\rho_{\text{s}}(t))=\sum_{n\neq n'}|\rho_{\text{s}}(t)[n,n']|$~\cite{Plenio2014}, can be attributed to the virtual processes involving simultaneous creation or annihilation of two excitations. These processes underscore the importance of the counter-rotating terms when considering non-Markovian effects. The temporal evolution of the coherence induced by the virtual processes, primarily driven by $\hat{\pmb{\rho}}_{\text{coh}}^{(\dagger)}$, exhibits oscillatory behavior [Fig. 3(c)]. {Remarkably, this cross-site coherence persists even beyond the relaxation time $t=\tau$ [Fig.~\ref{fig:4}(c), orange short vertical segment], even when strong on-site dephasing is present at every site~\cite{Lukin2008,Purkayastha2020}. This dephasing is modeled within the BMME framework [Eq.~(A24)~\cite{supplemental}] by incorporating the Lindblad dissipator $\sum_{n=1}^{N}\mathcal{J}(L_{\text{nos},n})$, where $L_{\text{nos},n}=\sqrt{\gamma_{\text{dep}}}d_n^{\dagger}d_n$.} 
This unexpected resilience arises from the interplay between the strong system-bath interactions and the dephasing. 
Through the virtual processes, the total system is effectively projected into the preferred eigenstates. These processes include not only the conventional transitions to 
$\approx |e_n,N_{\text{ph}}+1\rangle \pm |g_n,N_{\text{ph}}\rangle$, 
but also the higher-order transitions to 
$\approx |e_n,N_{\text{ph}}\rangle \pm |g_n,N_{\text{ph}}+1\rangle$. 
Here, $e_n$ ($g_n$) represents the occupied (vacant) particle state hybridized with either $N_{\text{ph}}+1$ or $N_{\text{ph}}$ from the bath. This hybridization maintains the coherence even in the presence of dephasing at every site~\cite{Lukin2008,Purkayastha2020,Filip2018}. These findings highlight the intriguing possibilities for the decoherence resistance within the regime of the non-Markovian skin effect.}

\begin{figure}[]
    \centering
    \includegraphics[width = 1\columnwidth]{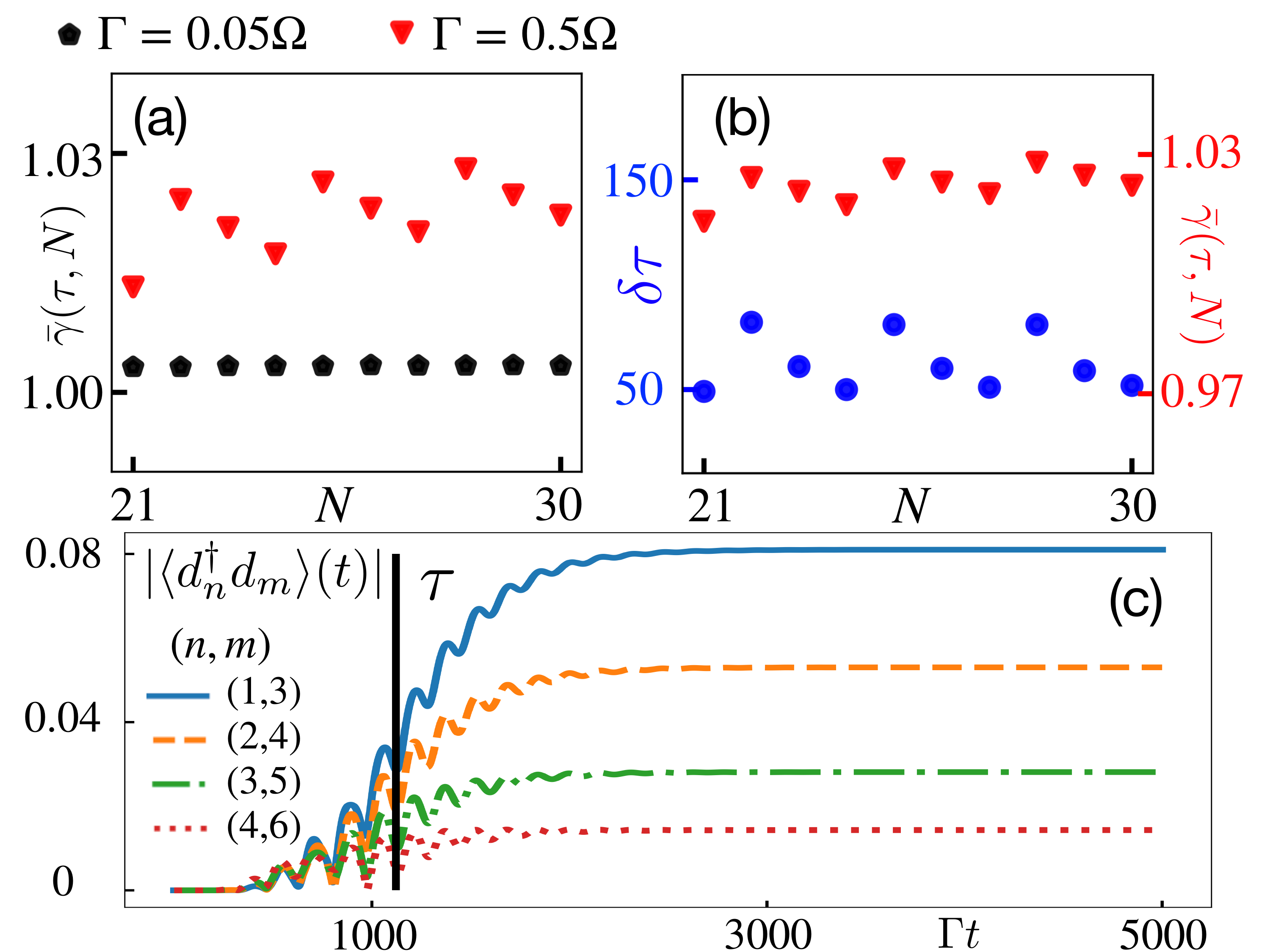}
    \caption{(a) $\Bar{\gamma}(\tau, N)$ versus $N$ for $\Gamma=0.05\Omega$ (black hexagons) and $\Gamma=0.5\Omega$ (red triangles). (b) Relaxation time deviation $\delta\tau$ between HEOM and BMME approaches (blue circles, left axis),and $\Bar{\gamma}(\tau, N)$ (red triangles, right axis), both as a function of $N$ for $\Gamma=0.5\Omega$. (c) Time evolutions of $|\langle d_{n}^{\dagger}d_{m}\rangle(t)|$ for different values of $(n,m)$. } 
    \label{fig4}
\end{figure}

\emph{Coherence delocalization}---We now stress an intriguing observation regarding the properties of the LSE in relation to $\tau$, which exhibits a linear dependence on $N$. However, what makes this behavior even more interesting is the presence of distinct oscillations in $\tau$ versus $N$. 
To delve into the underlying particle transport mechanisms driving this phenomenon, we calculate the average scaled rate of particle transport up to $\tau$ along this 1D chain~\cite{Superradiant2021}:
\begin{equation}
\begin{aligned}
\Bar{\gamma}(\tau,N)\approx \frac{1}{\tau}\sum_{n=1}^{N}\int_{0}^{\tau}\gamma(t,N)dt,
\label{eq:scaled_rate}
\end{aligned}
\end{equation}
{where $\gamma(t,N)=\sum_{n=1}^{N}[\langle d_{n}^{\dagger}d_{n}\rangle(t)+\sum_{n\neq m}\langle d_{n}^{\dagger}d_{m}\rangle(t)]$. The second term in $\gamma(t,N)$ accounts for superradiant contributions arising from inter-site coherence.} As the coupling strength $\Gamma$ approaches $0.5\Omega$ in the non-Markovian regime, we observe the increase in $\Bar{\gamma}(\tau, N)$ compared to the weak-coupling case as shown in Fig.~\ref{fig4}(a). This enhancement solely stems from the second term in Eq.~(\ref{eq:scaled_rate}), as the first term (representing individual site population) remains constant due to trace preservation~\cite{supplemental}. The intriguing small oscillations in the non-Markovian LSE $\tau$ observed in Fig.~\ref{fig1} correlate with sharp changes in the $\Bar{\gamma}(\tau, N)$ with respect to $N$. We quantify this connection by analyzing the deviation in $\tau$ obtained from HEOM compared to the BMME, i.e., $\delta\tau$. As shown in Fig.~\ref{fig4}(b), the periodic maxima and minima of $\delta\tau$ align with the corresponding peaks and valleys in $\Bar{\gamma}(\tau, N)$. {This alignment between $\delta\tau$ and $\Bar{\gamma}(\tau, N)$ can be understood through the analysis of the eigenmode energy spectrum~\cite{supplemental}.}.

{Note that the dynamics of $\gamma(t, N)$ constitute an important part of $\Bar{\gamma}(\tau, N)$. We observe that the $\langle d_{n}^{\dagger}d_{m}\rangle(t)$ term in $\gamma(t, N)$ clearly \textit{demonstrates the effect of virtual processes induced by the counter-rotating terms in the strong-coupling regime}. As illustrated in Fig.~\ref{fig4}(c), only cross-site $\langle d_{n}^{\dagger}d_{m}\rangle(t)$ terms contribute significantly to $\gamma(t, N)$, such as $(n,m)=(1,3)$, $(2,4)$, and even $(3,5)$, $(4,6)$. This is a phenomenon analogous to the cross-site coherence, i.e. no contribution from any two adjacent sites. An intuitive picture of this effective cross-site coupling, $\langle d_{n}^{\dagger}d_{m}\rangle(t)$, can be conceptualized using a three-site model: A particle moving from the third site to the second site, via the conventional transition operator $d_{n}^{\dagger}d_{n+1} a_k^{\dagger}$, generates a boson. This boson can subsequently undergo a virtual process (higher-order transition) facilitated by the counter-rotating term $d_{n}^{\dagger}d_{n+1} a_k$, enabling the particle to further transit to the first site. This composite transition process generates the cross-site $\langle d_{n}^{\dagger}d_{m}\rangle(t)$ terms, thereby creating the cross-site quantum correlations.}

In other words, the emergence of inter-site coherence, facilitated by the non-Markovian effect with virtual process, can hinder the skin effect's tendency to localize particles at the system edge. In this sense, the inter-site coherence plays a similar role to Anderson localization in the well-known competition between skin effect and Anderson localization in many-body systems. Both phenomena can impede the skin effect's localization tendency, influencing the particle transport dynamics~\cite{ChenShu2019}.

\emph{Conclusions}---Our investigation of the ``non-Markovian skin effect" reveals unique signatures not captured by conventional Markovian treatments. These include thick skin effect, persistent coherence in non-Markovian eigenmodes even under local decoherence, and coherence-delocalization effect at the relaxation time. These findings pave the way for exciting future research, such as exploring diverse systems and boundary conditions, delving into multi-particle scenarios, and investigating the impact on topological features like the winding number~\cite{Zhang2021}. Such investigations hold promise for unlocking novel applications in various areas, including controlling light~\cite{Longhi2015,Weidemann2020} and electron localization~\cite{ChenShu2019,Jiangbin2020,Xing2023}, manipulating directional transport phenomena~\cite{2Longhi2015,YanBo2020,Liew2020,Ghaemi2021}, influencing critical entanglement properties~\cite{Shinsei2023}, and enhancing the sensitivity of topological sensors~\cite{Bergholtz2020,Wiersig2014}.

\emph{Acknowledgment}---F.N. is supported in part by:
Nippon Telegraph and Telephone Corporation (NTT) Research,
the Japan Science and Technology Agency (JST)
[via the CREST Quantum Frontiers program Grant No. 24031662, the Quantum Leap Flagship Program (Q-LEAP), and the Moonshot R\&D Grant Number JPMJMS2061], 
and the Office of Naval Research Global (ONR) (via Grant No. N62909-23-1-2074). YNC acknowledges the support of the National Center for Theoretical Sciences and the National Science and Technology Council, Taiwan (NSTC Grants No. 112-2123-M-006-001).

%

\end{document}